\documentclass[12pt]{iopart}

\usepackage[all]{xy}
\usepackage{mathrsfs}
\usepackage{graphicx}% Include figure files
\usepackage{epstopdf}
\usepackage{dcolumn}% Align table columns on decimal point
\usepackage{bm}% bold math
\usepackage{color} % change color of text
\usepackage{CJK}
\newcommand{\beq}{\begin{equation}}
\newcommand{\eeq}{\end{equation}}
\newcommand{\beqa}{\begin{eqnarray}}
\newcommand{\eeqa}{\end{eqnarray}}

\begin{document}
\title{Shortcuts to adiabatic rotation of a two-ion chain}
\author{Ander Tobalina}
\address{Departament of Physical Chemistry, University of the Basque Country (UPV/EHU), Apdo.
644, 48080 Bilbao, Spain}
\author{Juan Gonzalo Muga}
\address{Departament of Physical Chemistry, University of the Basque Country (UPV/EHU), Apdo.
644, 48080 Bilbao, Spain}
\author{Ion Lizuain}
\address{Department of Applied Mathematics, University of the Basque Country UPV/EHU, Donostia-San Sebastian, Spain}
\author{Mikel Palmero}
\ead{mikel.palmero@ehu.eus}
\address{Department of Applied Physics, University of the Basque Country (UPV/EHU), 48013 Bilbao, Spain}
\begin{abstract}
We inverse engineer fast rotations of a linear trap with two ions
for a predetermined rotation angle and time, avoiding final excitation.
Different approaches are analyzed and compared  when the ions are of the same species or of different species.
%For different species there is no point transformation to find independent modes.
The separability into dynamical normal modes for equal ions in a common harmonic trap, or for different ions
in non-harmonic traps with up to quartic terms allows for simpler computations of the rotation protocols.
For non-separable scenarios, in particular
for different ions in a harmonic trap, rotation protocols are also found using more costly numerical optimisations.
\end{abstract}

%\maketitle

\section{Introduction}
%
%
%
%\tr{A major challenge in modern atomic physics is to develop a scalable architecture for quantum information
%processing. A proposed scheme to achieve scalability is based on shuttling individual ions or small groups of ions
%among processing or storing sites  \cite{Kielpinski2002,Rowe2002,Reichle2006,Home2009,Roos2012,Monroe2013}.
%Apart from shuttling \cite{Torrontegui2011,Palmero2013,Palmero2014,Lu2014,Lu2015},  other manipulations of the ion motion would be needed,
%such as expansions or compressions of the trap \cite{Chen2010,Palmero2015a}, separating or merging ion chains \cite{Lau2012,Palmero2015,Kaufmann2014}, and rotations \cite{Splatt2009}.
%All these basic dynamical operations should fulfill two seemingly contradictory requirements:
%they should be fast, but free from final motional excitation. Shortcuts to adiabaticity for ``fast and safe driving''
%have been proposed for several of these elementary operations \cite{Torrontegui2011,Palmero2013,Palmero2014,Lu2014,Lu2015,Chen2010,Palmero2015a,Palmero2015} and have also been implemented experimentally \cite{Bowler2012,Walther2012}.}

%Among the quantum systems that we can prepare,  drive and measure ,
Trapped ions stand out as a flexible architecture to control
internal and/or  motional states and dynamics
%, as well mixed operations with motional-internal couplings,
for fundamental research
of quantum phenomena and technological applications.
Pure motional control without internal state transitions is in particular crucial in proposals of two-qubit gates, see e.g.  \cite{Palmero2017}, or interferometry
\cite{Campbell2017,Martinez-Garaot2018,Rodriguez-Prieto2020}, as well as to scale up the number of ions for quantum information processing \cite{Kielpinski2002,Rowe2002,Reichle2006,Home2009,Roos2012,Monroe2013,Kaushal2020,Wan2020}.
The toolbox of basic operations induced by controlling the voltage of electrodes  in different Paul trap configurations or detuned laser fields
includes transport, expansions and compressions, separation and merging
of ion chains, and rotations, the latter being the central topic of this work.

Specific motivations to implement
rotations are: reordering an ion chain (to scale up quantum information processing
or to locate  cooling ions at appropriate positions) \cite{Splatt2009,Kaufmann2017};
rotation sensing \cite{Campbell2017};
different simulations (e.g. of black holes \cite{horstmann2010} or diatomic molecules \cite{Urban2019});
probing the exchange phase of quantum statistics
 \cite{roos2017};  or  sorting ions  according to charge and mass \cite{Masuda2015}.

Trap rotations, to impart some angular momentum to an ion or ion chain, or to
reorient the
longitudinal axis of the trap,  have been implemented
in experiments with improving accuracy \cite{Splatt2009,Urban2019,Kaufmann2017,vanMourik2020}, and investigated theoretically \cite{Palmero2016,Lizuain2017}.

Motional control operations, and rotations in particular, need in most applications to be fast, relative to adiabatic dynamics,  but also gentle, avoiding final excitations,
two requirements met with shortcut to adiabaticity (STA) driving protocols \cite{Guery2019}.
There are different STA techniques but,  for trapped ion driving, STA invariant-based inverse engineering has proven useful \cite{Palmero2017, Martinez-Garaot2018, Rodriguez-Prieto2020, Lizuain2017, Torrontegui2011, Palmero2013, Palmero2014, Lu2014, Lu2015, Chen2010, Palmero2015a, Palmero2015, Sagesser2020},
also to design trap rotations for a single ion \cite{Palmero2016}.

In this paper we extend  to a two-ion chain the  design of STA 1D-trap rotations  done in reference \cite{Palmero2016}.
Our aim is to  inverse engineer the rotation angle  to implement a fast  process, free from final excitations.
The work in reference \cite{Palmero2016}  was indeed presented  as a preliminary step towards the more complex scenario of the chain rotation,
which allows for different, and surely more relevant applications, in particular reordering.
Engineering the two-ion rotation also entails non-trivial technical complications
due to the increase in the number of equations to be solved, and also because, for some configurations, in particular
for two different species in a harmonic trap,
there is not in  general a  point-transformation that provides independent dynamical normal modes \cite{Lizuain2017}\footnote{``Dynamical normal modes" generalise regular (static) normal modes. They are independent concerted motions represented by harmonic oscillators
with time-dependent parameters \cite{Lizuain2017, Palmero2014}, generally with a time-dependent oscillation frequency.}.
Inverse engineering is much easier -to describe the motion and with respect to computational time-
for independent modes than for a system which is not separable by point-transformations.\footnote{Separability by non-point transformations is possible in principle but it is considerably more involved in terms of its interpretation and practical use.
Its application  to inverse engineer one-particle rotations in anysotropic traps was explored in \cite{Lizuain2019} under  some strong restrictions in process timing and rotation speed.}
%These complications deserve a specific study per se, and

%We assume that  the ions are trapped in a linear, harmonic trap, tightly confined in a radial
%direction so that they move effectively along a one-dimensional axial direction, hereafter ``the line''.
%We also assume the ions to never change the ordering within the trap frame, due to the strong Coulomb repulsion.

% with respect to a
%vertical axis that crosses the centre of the trap.
%Such an operation was adiabatically performed by Splatt et al \cite{Splatt2009} where the objective was simply showing the reordering of small ion chains was possible. This is important, for example, when topologically encoding a qubit
%\cite{Nigg2014},
%where the ion chain reordering is essential.
%
%
%
% % % % % % % % % % % % % % % % % % % % % % % % % % % % % % % % % % % % % % % % %
% % % % % % % % % % % % % % % % % % % % % % % % % % % % % % % % % % % % % % % %
\begin{figure}[t]
\begin{center}
\includegraphics[width = 0.6 \linewidth]{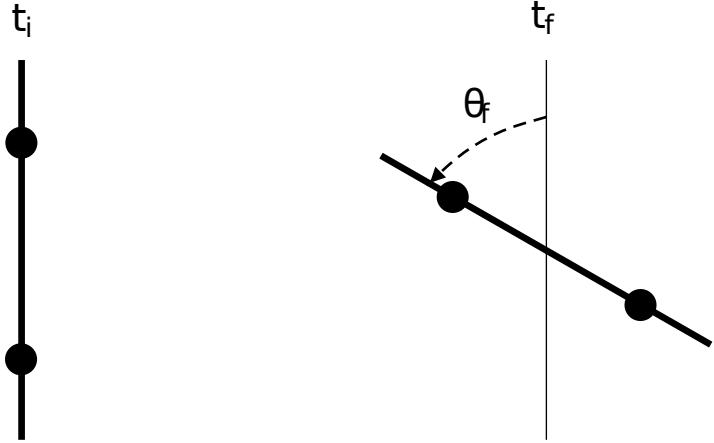}
\caption{\label{fig7_1}(Color online)
Schematic representation of the rotation of two ions confined along a line rotated by an angle $\theta_f$
in a time time $t_f$.}
\end{center}
\end{figure}
% % % % % % % % % % % % % % % % % % % % % % % % % % % % % % % % % % % % % % % % % % %
% % % % % % % % % % % % % % % % % % % % % % % % % % % % % % % % % % % % % % % % % % %
%
%
%In this section, we provide the Hamiltonian and the notation that describes the physical setting.
%As the ion's motion is constrained to a rotating, one-dimensional, horizontal  line,

We introduce now the basic model. We opt for a cavalier, idealised modelling where
the trap is assumed for simplicity to be tightly confined in the radial
direction, as depicted in figure \ref{fig7_1}, i.e., we leave aside peculiarities of the
experimental settings, such as micromotion effects and detailed electrode configurations, that may vary significantly among different traps.
Our solutions will therefore be guiding starting points for a realistic implementation
%that will need
%to take into account filtering, calibration, 3D effects, and other sources of imperfect implementation of the ideal potentials
\cite{vanMourik2020, Kaufmann2018}.

The trapping line rotates in a horizontal plane in a time $t_f$ up to a predetermined final angle,  $\theta_f= \pi$ in all examples.
{We first find
the classical Hamiltonian from the corresponding  Lagrangian and then quantise the result.
%(The ions may safely be treated as distinguishable because the probability density clouds
%do not overlap due to the strong Coulomb repulsion.)
Let $s_i$, $i=1,2$,  denote the points on the line where each ion lays. $s_i$
may take positive and negative values.  The Cartesian (lab frame) components of a trajectory $s_i(t)$ are
$x_i=x_i(s,t)$,
$y_i=y_i(s,t)$,
\beq
x_i = s_i\cos (\theta), \quad y_i=s_i\sin (\theta),
\eeq
where $\theta=\theta (t)$ is the rotation angle.}
For two different ions in a common trap potential $f(s_i)$ the Lagrangian is (we have considered that
the  magnetic interaction between the two moving charges can be safely neglected, see Appendix A)
\beqa
\label{Lagrangian_2diffion}
L = \sum_{i=1,2}\left[\frac{m_i}{2} \dot{s}_i^2-f(s_i)+ \frac{m_i{\dot\theta}^2}{2}s_i^2\right]-\frac{C_c}{s_2-s_1},
\eeqa
%\omega_a^2 &=& \omega_1^2 -\dot{\theta}^2,\quad
%\nonumber\\
%where
%
%\beq
%\quad u_i= m_i(\omega_i^2 -\dot{\theta}^2), \quad m_i\omega_i^2=k,
%\eeq
%
with corresponding
Hamiltonian
%The kinetic energy  is $K=\frac{1}{2}m_1(\dot{x}_1^2+\dot{y}_1^2)+\frac{1}{2}m_2(\dot{x}_2^2+\dot{y}_2^2)$ and the potential energy
%is assumed by now to be harmonic plus a Coulomb interaction between both ions, $V=u_0 (s_1^2+s_2^2)/2+{C_c}/{(s_2-s_1)}$,
%where $u_0=m\omega_0^2$ and $C_c={e^2}/{(4\pi\epsilon_0)}$, $\epsilon_0$ being the vacuum permittivity, $e$ the electric charge of a single electron,
%$\omega_0$ the external harmonic frequency for both ions
%in the  direction of the line, and $m$ the mass of each ion.
%The Lagrangian takes the form
%
%
%\beqa
%\label{Lagrangian_2eqion}
%L &=& \frac{1}{2} m\dot{s}_1^2+\frac{1}{2}m\dot{s}_2^2-\frac{1}{2}m\omega^2 (s_1^2+s_2^2)-\frac{C_c}{s_2-s_1},
%\\
%\omega^2 &=& \omega_0^2 -\dot{\theta}^2.
%\label{ome_ef}
%\eeqa
%
%
%\section{General Hamiltonian}
%
%Consider the following Hamiltonian (in principle we consider different masses),
%
\begin{eqnarray}
H&=&\frac{p_1^2}{2m_1}+\frac{p_2^2}{2m_2}+V,
\label{Hgeneral}\\
V&=&\sum_{i=1,2}\left[f(s_i)-\frac{m_i}{2}{\dot{\theta}}^2 s_i^2 \right]+\frac{C_c}{s_2-s_1}.
\label{pot}
\end{eqnarray}
In the Coulomb repulsion term
%$u_i=m_i (\omega_i^2-\dot\theta^2)$,
$C_c={e^2}/{(4\pi\epsilon_0)}$, where $\epsilon_0$ is the vacuum permittivity, and $e$ the electric charge of the electron.
%%%%%%%%%%%%%%%%%%%%%%%%%%%%%%%%%%%%%%%%%%%%%%%%%%%%%%%%%
%
%\section{Harmonic approximation}

%\begin{itemize}
The equilibrium positions $\{s_i^{(0)}\}$ of the ions
are found by solving the set of equations $\{{\partial V}/{\partial s_i}=0\}$.
Since different external traps may be considered, the following equations are for a generic $V$, the results for the simple harmonic trap
are given later in section \ref{htrap}.

We define the equilibrium distance between ions as
\beq
d=s_2^{(0)}-s_1^{(0)}
%=\left[\frac{C_c(u_1+u_2)}{u_1u_2}\right]^{1/3},
\eeq
and
expand $V$ around the equilibrium positions, keeping terms up to second order. Using mass-weighted coordinates
$\tilde s_i=\sqrt{m_i} s_i$ and
momenta $\tilde p_i=p_i / \sqrt{m_i}$, $H$ is simplified to the quadratic form
%
% \begin{eqnarray}
% H&=&\frac{\tilde p_1^2}{2}+\frac{\tilde p_2^2}{2}+(\tilde s_1-s_1^{(0)},\tilde s_2-s_2^{(0)})
%
% \nonumber\\
%&& \left(\begin{matrix}
%\frac{\frac{2 C_c}{d^3}+u_1}{m_1} & -\frac{2 C_c}{d^3\sqrt{m_1 m_2} }   \\
%-\frac{2 C_c}{d^3\sqrt{m_1 m_2}} & \frac{\frac{2 C_c}{d^3}+u_2}{m_2}
%       \end{matrix}\right)
%
% \left(\begin{matrix}\tilde s_1-s_1^{(0)}\\ \tilde s_2-s_2^{(0)} \end{matrix}\right).
%\end{eqnarray}
%
%This structure is in fact valid generally in the small-oscillation regime
%for more general external potentials, as the ones used in Appendix B,
%!!!!!!!!!!FIX THIS
\begin{equation}
\label{general}
 H=\frac{\tilde p_1^2}{2} + \frac{\tilde p_2^2}{2}
 %\nonumber\\
 \!+\!(\tilde s_1-s_1^{(0)},\tilde s_2-s_2^{(0)}) {{\bm{\mathsf{ v}}}}
%\left(\begin{matrix}
%v_{11}& v_{12}   \\
%v_{21}& v_{22}
%       \end{matrix}\right)
 %
 \left(\begin{array}{l}
   \tilde{s}_1-s_1^{(0)}\\
   \tilde{s}_2-s_2^{(0)}
 \end{array}\right),
\end{equation}
where the matrix $\bm{\mathsf v}$ has elements ${\sf v}_{ij}=\frac{1}{\sqrt{m_im_j}}\left.\frac{\partial^2 V}{\partial s_i\partial s_j}\right|_{\{s_i,s_j\}=\{s_i^{(0)},s_j^{(0)}\}}$.
%and the $s_i^{(0)}$ are found from the potential minimum condition.

%\item  Con este Hamiltoniano (ya que tenemos un coupling tipo $s_1s_2$), podemos explorar la opci\'on de los invariantes 2D
%delPRA2020 \cite{Tobalina2020}. Sospechamos que solo podemos conseguir resultados parciales
%(que se conserve la energ\'ia de un modeo por ejemplo, pero no la total).
%
%\end{itemize}

%%%%%%%%%%%%%%%%%%%%%%%%%%%%%%%%%%%%%%%%%%%%%%%%%%%%%%%%%
\section{Diagonalisation and dynamical normal modes: setting the equations}

We may try to decouple the dynamics by diagonalising ${\bm{\mathsf{v}}}$. As explained in reference \cite{Lizuain2017},
moving to a frame defined by the eigenvectors of ${\bm{\mathsf{v}}}$ leads,
after a classical canonical transformation or, equivalently, quantum unitary transformation, to the following effective Hamiltonian
\cite{Lizuain2017}
\begin{eqnarray}
 H'=\sum_{\nu=\pm}\left[\frac{p_\nu^2}{2}+\frac{\Omega_\nu^2}{2}\left(s_\nu+\frac{\dot p_{0\nu}}{\Omega_\nu^2}\right)^{\!\!2}\right]\!\!-\!\dot \mu (s_-p_+-s_+p_-),
\label{effective}
\end{eqnarray}
%solving the set of
%
where $\mu$ is the tilting angle of the potential in configuration space defined by the relation
\begin{eqnarray}
\label{decoupling_cond}
\tan 2\mu=\frac{2v_{12}}{v_{11}-v_{22}},
%=\frac{4 C_c \sqrt{m_1m_2}}{(m_1-m_2) \left(2 C_c+d^3 k\right)},
\end{eqnarray}
%
%where the second equality is for the harmonic trap in equation (\ref{Hamiltonian_2ionrot}),
and the momentum shifts
\begin{eqnarray}
% \left\{\begin{matrix}
 p_{0+}&=&\dot s_1^{(0)} \sqrt{m_1}\cos\mu +\dot s_2^{(0)}\sqrt{m_2}\sin\mu,
 \\
 p_{0-}&=&-\dot s_1^{(0)} \sqrt{m_1}\sin\mu +\dot s_2^{(0)} \sqrt{m_2}\cos\mu,
% \end{matrix}\right.
\end{eqnarray}
have been defined.
The  coordinates that diagonalise ${\bm{\mathsf{v}}}$ are
 \begin{eqnarray}
%\left\{\begin{matrix}
\!\!\!\!\!\!\!s_+&=&\sqrt{m_1} (s_1\!-\! s_1^{(0)}) \cos\mu\! +\! \sqrt{m_2} (s_2\!-\!s_2^{(0)})\sin\mu,\\
\!\!\!\!\!\!\!s_-&=&-\sqrt{m_1} (s_1\!-\!s_1^{(0)}) \sin\mu\! +\! \sqrt{m_2} (s_2\!-\!s_2^{(0)}) \cos\mu,
%\end{matrix}\right.
\end{eqnarray}
with conjugate momenta
 \begin{eqnarray}
%\left\{\begin{matrix}
p_+&=&\frac{\cos\mu}{\sqrt{m_1}}p_1 + \frac{\sin\mu}{\sqrt{m_2}}p_2,\\
p_-&=&-\frac{\sin\mu}{\sqrt{m_1}}p_1 + \frac{\cos\mu}{\sqrt{m_2}}p_2.
%\end{matrix}\right.
\end{eqnarray}
The squares of the frequencies are
\begin{eqnarray}
%\left\{\begin{matrix}
 \Omega_+^2&=&v_{11}\cos^2\mu+v_{22}\sin^2\mu+v_{12}\sin{2\mu},
 \nonumber\\
 \Omega_-^2&=&v_{11}\sin^2\mu+v_{22}\cos^2\mu-v_{12}\sin{2\mu}.
% \end{matrix}\right.
\label{omegas}
\end{eqnarray}
$s_\pm$ describe independent, dynamical  normal modes whenever $\mu$ is time independent, see equation (\ref{effective}). In a quantum scenario this means that
any wave-function dynamics can be decomposed in terms of the dynamics of two independent
harmonic oscillators with time-dependent parameters. Different scenarios to achieve this decoupling are considered
in the following.
\subsection{Results for the harmonic trap\label{htrap}}
For the common harmonic external potential with spring constant $k$,
$f(s_i)={k}s_i^2/2$, which is
the only configuration considered hereafter in the main text, the potential (\ref{pot}) takes the form
\begin{eqnarray}
V&=&\frac{1}{2}u_1 s_1^2+\frac{1}{2}u_2 s_2^2+\frac{C_c}{s_2-s_1},
\label{Hamiltonian_2ionrot}
\end{eqnarray}
where
\beq
u_i= m_i(\omega_i^2 -\dot{\theta}^2), \quad m_i\omega_i^2=k.
\label{spco}
\eeq
The $u_i$ are effective spring constants affected by the rotation speed.
Unless $m_1=m_2$, they are different for both ions.  With this $V$ we find the explicit relations

\begin{eqnarray}
  s_i^{(0)}&=&-\left[\frac{C_c u_j^2}{u_i (u_i+u_j)^2}\right]^{\!1/3}\!, i\ne j,
   %\nonumber\\
  %s_2^{(0)}=\left[\frac{C_c u_1^2}{u_2 (u_1+u_2)^2}\right]^{\!1/3}\!.
  %\end{eqnarray}
  %
  \nonumber\\
  d&=&\left[\frac{C_c(u_1+u_2)}{u_1u_2}\right]^{1/3},
  \nonumber\\
  {\bm{\mathsf v}}&=&\left(\begin{array}{lr}
      \frac{\frac{2 C_c}{d^3}+u_1}{m_1} & -\frac{2 C_c}{d^3\sqrt{m_1 m_2} }   \\
      -\frac{2 C_c}{d^3\sqrt{m_1 m_2}} & \frac{\frac{2 C_c}{d^3}+u_2}{m_2}
   \end{array}\right),
  \nonumber\\
  \label{decoupling_condh}
  \tan 2\mu&=&\frac{4 C_c \sqrt{m_1m_2}}{(m_1-m_2) \left(2 C_c+d^3 k\right)}.
\end{eqnarray}
\section{Equal ions}
If the ions are equal, $m_1=m_2=m$, the tilting angle takes the constant value $\mu=-\pi/4$.
The decoupling condition  is therefore identically satisfied at all times.
% without further
%assumptions on the value of $\dot\theta$.
Also,  $u_1=u_2=u=m\omega^2$, with
\beq
\omega^2=\omega_0^2-\dot\theta^2
\label{ome_ef}
\eeq
and $\omega_0$ constant. The angular velocity of the rotation $\dot{\theta}(t)$    could be negative at some intervals, whereas $\omega^2$
may also be positive or negative.
%,  with  $\omega$ purely imaginary if $\omega^2<0$.
The equilibrium positions are simplified to
\begin{eqnarray}
s_2^{(0)}=-s_1^{(0)}=\frac{x_0}{2}\quad\textrm{with} \quad x_0=\left(\frac{2Cc}{m\omega^2}\right)^{1/3},
\end{eqnarray}
which are symmetrical with respect to the trap centre $s=0$.
The decoupled, effective Hamiltonian is therefore
\begin{eqnarray}
 H''=\sum_{\nu=\pm}\left[\frac{p_\nu^2}{2}+\frac{\Omega_\nu^2}{2}\left(s_\nu+\frac{\dot p_{0\nu}}{\Omega_\nu^2}\right)^2\right],
\label{18}
\end{eqnarray}
with
\begin{eqnarray}
%\left\{\begin{matrix}
s_\pm&=&\sqrt{\frac{m}{2}}\left[\left(s_1+\frac{x_0}{2}\right)\mp\left(s_2-\frac{x_0}{2}\right)\right],
\nonumber\\
p_{0+}&=&- \sqrt{\frac{m}{2}} \dot x_0,
%
%s_-&=&\sqrt{\frac{m}{2}}\left[(s_1+\frac{x_0}{2})+(s_2-\frac{x_0}{2})\right],
\quad p_{0-}=0,
\nonumber\\
\Omega_+^2&=&3\omega^2,\quad \Omega_-^2=\omega^2.
\label{OmegaNM}
%\end{matrix}\right.
\end{eqnarray}

%\subsection{Different ions I: $2 C_c+d^3 k=0$}
%If the masses are different, we could think in forcing $2 C_c+d^3 k=0$. (*) Esto es lo que est\'a ahora en el manuscrito:
%``which in order to be $d$ real requires a repulsive potential'' (...). Entiendo que sale una $d$  negativa, y eso nos da problemas. Por
%ejemplo el potencial (ver la matriz) dejar\'ia de ser atractivo para convertirse en repulsivo (.....)

%%%%%%%%%%%%%%%%%%%%%%%%%%%%%%%%%%%%%%%%%%%%%%%%
%\section{Different ions, beyond harmonic potential}
%Up to now everything has been harmonic. Hamiltonian only has quadratic terms + Coulomb interaction. Different proposals so far:
%(i) linear term, (ii) double well potential (iii) cubic term, (iv) double well+ cubic term, etc...

We consider rotation protocols with a smooth behavior of $\theta$ at the
boundary times $t_b=0,t_f$,
\beqa
\theta(0)&=&0,\quad \theta(t_f)=\theta_f,
\label{cond1_2ion}\\
\dot{\theta}(t_b)&=&\ddot{\theta}(t_b)=0.
\label{cond2_2ion}
\eeqa
These conditions imply that
\beqa
\omega(t_b)&=&\omega_0,
\\
\dot\omega(t_b)&=&\ddot\omega(t_b)=\dot{p}_{0\nu}(t_b)=0.
\label{cond3_2ion}
\eeqa
%
%The Hamiltonian is uncoupled,
%
%\begin{eqnarray}
% H'=\sum_{\nu=\pm}\left[\frac{p_\nu^2}{2}+\frac{\Omega_\nu^2}{2}\left(s_\nu+\frac{\dot p_{0\nu}}{\Omega_\nu^2}\right)^{\!\!2}\right]
%\end{eqnarray}
%
%The system is formally akin to two coupled harmonic oscillators with the same time-dependent frequency, a system with no known Lewis-Riesenfeld invariants \cite{Lewis1969}. Recently, a different type of invariant has been used to implement a shortcut on two coupled harmonic oscillators \cite{Tobalina2020}. Here, however, we define the
%normal modes to get an approximate Hamiltonian, allowing ourselves to apply the usual tools of inverse engineering through dynamical invariants, as we have done in previous works \cite{Palmero2014,Palmero2015a,Palmero2015}.

%
% % % % % % % % % % % % % % % % % % % % % % % % % % % % % % % % % % % % % % % % %
% % % % % % % % % % % % % % % % % % % % % % % % % % % % % % % % % % % % % % % %
\begin{figure*}[t]
\begin{center}
\includegraphics[width=0.4 \linewidth]{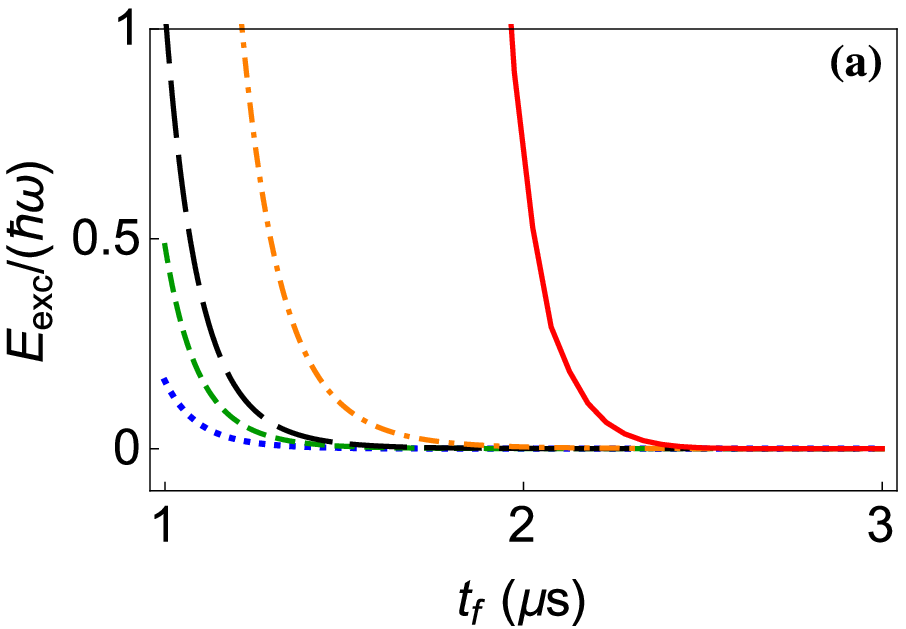}
\includegraphics[width=0.42 \linewidth]{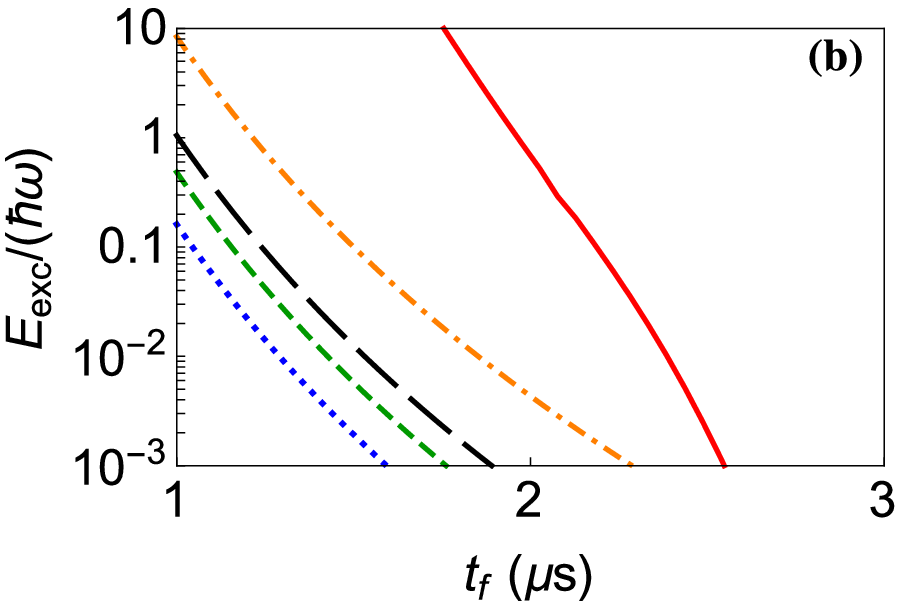}
\caption{\label{fig7_2}(Color online)
Two equal ions. Exact energy excess (final minus initial energy) starting from the ground state and with dynamics driven by the full potential (\ref{Hamiltonian_2ionrot}) according to equation (\ref{controlparameter}) for the parameters $c_{3-6}$ that minimise the excitation in the normal modes.
(a) represents this excitation in a linear scale and (b) in a logarithmic scale.
Dotted blue line:  protocol using all 4 free parameters in the ansatz for $\theta(t)$;
Short-dashed green line: only 3 free parameters, i.e., $c_6=0$;
Long-dashed black line: only 2 free parameters,  i.e., $c_5=c_6=0$;
Dash-dotted orange line: only one free parameter, i.e., $c_4=c_5=c_6=0$;
Solid red line fixes: $c_{3-6}=0$ so there is no optimisation.
The evolution was done for two $^{40}$Ca$^+$ ions, with an external trap frequency $\omega_0/(2\pi)=1.41$ MHz and a total rotation angle $\theta_f=\pi$.
}
\end{center}
\end{figure*}
% % % % % % % % % % % % % % % % % % % % % % % % % % % % % % % % % % % % % % % % % % %
% % % % % % % % % % % % % % % % % % % % % % % % % % % % % % % % % % % % % % % % % % %

The two independent harmonic oscillators expand or compress through the time dependence of $\Omega_\nu$ and experiment a ``transport'',
in $s_\nu$ space, along $\dot{p}_{0\nu}/\Omega_\nu^2$.
%The Hamiltonian is similar to the one  in \cite{Palmero2015a} for simple trap expansions or compressions, except for the effective lab frequency $\omega$ having the dependence given in equation~\ref{ome_ef}, while in \cite{Palmero2015a} $\omega$ was the time dependent frequency of the external trap, which was directly controlled in the lab. {Here, its time dependence is due to $\theta(t)$, which is our control parameter. That means that it is not enough to design an $\omega$ that leaves the normal modes unexcited as done in \cite{Palmero2015a}. We additionally need to inverse engineer this parameter to get an expression of $\theta$ that still satisfies the conditions in Eqs.~\ref{cond1_2ion} and~\ref{cond2_2ion}.} This inverse engineering already proved to be problematic in \cite{Palmero2016} for a single ion rotation, because it implies a square root, $\dot{\theta}=\sqrt{\omega_0^2-\omega^2}$, which can be imaginary if the effective frequency $\omega$ (which, in principle, has no physical limitation for not being a ``real'' frequency) happens to be larger than the external trap frequency $\omega_0$.
The Hamiltonian~(\ref{18}) has a dynamical invariant \cite{Lewis1969}
\begin{equation}
I = \sum_{\nu=\pm} \frac{1}{2}[b_\nu(p_\nu-\dot{\alpha}_\nu)-\dot{b}_\nu(s_\nu-\alpha_\nu)]^2
+ \frac{1}{2}\Omega_{0\nu}^2\left(\frac{s_\nu-\alpha_\nu}{b_\nu}\right)^2,
\end{equation}
where $\Omega_{0\pm}=\Omega_\pm(0)$, and $b_\pm$ (scaling factors of the normal mode wavefunctions) and $\alpha_\pm$ (reference classical trajectories
for each oscillator) are auxiliary functions that have to satisfy, respectively, the Ermakov and Newton equations,
\beqa
\label{auxiliaryequations}
\ddot{b}_\pm+\Omega_\pm^2 b_\pm &=& \frac{\Omega_{0\pm}^2}{b_\pm^3},
\\
\ddot{\alpha}_\pm+\Omega_\pm^2\alpha_\pm &=& \dot{p}_{0\pm}.
\eeqa
The time-dependent Schr\" odinger equation can be solved
by superposing, with constant coefficients,  elementary solutions which are also eigenstates of the invariant, with the (``Lewis-Riesenfeld'') phase adjusted to be also solutions of the  Schr\"odinger equation \cite{Torrontegui2011},
\beq
|\psi_{n\pm}''\rangle = e^{\frac{i}{\hbar}\left[\frac{\dot{b}_\pm s_\pm^2}{2b_\pm}+(\dot{\alpha}_\pm b_\pm-\alpha_\pm\dot{b}_\pm)\frac{s_\pm}{b_\pm}\right]}\frac{1}{\sqrt{b_\pm}}\Phi_n(\sigma_\pm),
\label{elementary}
\eeq
where $\sigma_\pm=\frac{s_\pm-\alpha_\pm}{b_\pm}$ and $\Phi_n$ are the eigenfunctions for the static harmonic oscillators with frequencies $\Omega_{0,\pm}$. The average energies for the $n$th elementary solution of each mode can be calculated  \cite{Palmero2015a, Palmero2015},
\beqa
\label{EnergyNM}
E''_{n\pm} &=& \langle \psi''_{n\pm}|H''|\psi''_{n\pm}\rangle\nonumber\\
                 &=& \frac{(2n+1)\hbar}{4\Omega_{0\pm}}\left(\dot{b}^2_\pm+\Omega_\pm^2b_\pm^2+\frac{\Omega_{0\pm}^2}{b_\pm^2}\right)
                 + \frac{1}{2}\dot{\alpha}_\pm^2+\frac{1}{2}\Omega_\pm^2\left(\alpha_\pm-\frac{\dot{p}_{0\pm}}{\Omega_{\pm}^2}\right)^2.
\eeqa
As $\dot{p}_{0\nu}(t_f)=0$, the final values are minimised when  the only contribution is due to the eigenenergies for the oscillators, with
\beqa
b_\pm(t_f)&=&1,\alpha(t_f)=\dot{\alpha}(t_f)=\dot{b}_\pm(t_f)=0.
\label{bco}
%
%\alpha(t_f)&=&\dot {p}_{0\pm}/\Omega_{\pm}^2|_{t_f}
\eeqa
\subsection{Inverse engineering}
\label{equalionssec}
Imposing commutativity between Hamiltonian and invariant at initial $t=0$ and final times $t=t_f$, the invariant drives the initial eigenstates of $H$ to corresponding final eigenstates along the elementary solutions (\ref{elementary}), although there could be diabatic excitations at intermediate times, when the commutation between Hamiltonian and invariant is not guaranteed. By inspection of equation (\ref{elementary}), commutativity at the boundary times
is achieved if the conditions in equation (\ref{bco}) are satisfied,
%
%\beqa
%\label{boundary1_2ion}
%\alpha_\pm(t_b)&=&0,\quad b_\pm(t_b)=1,
%\nonumber\\
%\dot{b}_\pm(t_b) &=&\dot{\alpha}_\pm(t_b)=0,
%\nonumber\\
%\dot{\alpha}_\pm(t_b) &=&0,
%\eeqa
%
%where $t_b=0, t_f$. As the external trap has the same frequencies at both time boundaries, $b_\pm(t_b) = 1$. We also impose
%
which occur automatically when the final energies (\ref{EnergyNM}) are minimised.
%\beqa
%\label{boundary2_2ion}
%,\quad
%\nonumber\\
%\alpha_\pm(t_b)=\ddot{\alpha}(t_b)=0
%\eeqa
%
%so the forcing term $\dot{p}_{0\pm}$ must vanish at the boundaries.
%\tr{The usual procedure is to design the normal mode dynamics, by trying ansatzes for $b_\pm$ and $\alpha_\pm$ that satisfy the boundary conditions in Eqs.~\ref{boundary1_2ion} and~\ref{boundary2_2ion} and then inversely obtain the control parameter(s). In this case, the control parameters are the rotation angle $\theta$ and the external trap frequency $\omega_0$, although we will preferably leave $\omega_0$ constant for a rigid rotation. However, as we said in the previous section, the inverse engineering of the rotation angle is problematic as it implies a square root. }
To inverse engineer the rotation we proceed similarly to reference   \cite{Palmero2016}, with an ansatz for $\theta(t)$ that satisfies boundary conditions~(\ref{cond1_2ion}) and~(\ref{cond2_2ion}) with some free parameters.
We
use up to 4 free parameters,
\beqa
\label{controlparameter}
\theta (t) &=& \frac{1}{16}(32c_3+80c_4+144c_5+224c_6-9\theta_f)\cos\left(\frac{\pi t}{t_f}\right)
\nonumber\\
               &-& \frac{1}{16}(48c_3 + 96 c_4 + 160 c_5 + 240 c_6 - \theta_f)\cos\left(\frac{3\pi t}{t_f}\right)
\nonumber\\
               &+& c_3\cos\left(\frac{5 \pi t}{t_f}\right) + c_4\cos\left(\frac{7 \pi t}{t_f}\right)
+ c_5 \cos\left(\frac{9 \pi t}{t_f}\right) + c_6 \cos\left(\frac{11 \pi t}{t_f}\right) + \frac{\theta_f}{2}.\nonumber\\
\eeqa
This gives an expression of $\dot{\theta}$, from which we find $\omega$ in equation~(\ref{ome_ef}). We introduce $\omega$ in~(\ref{OmegaNM}) to get the normal mode angular frequencies $\Omega_\pm$ needed in  the Ermakov equation~(\ref{auxiliaryequations}).
For a given set of values of these parameters  we solve the ``direct problem'' (Ermakov and Newton equations)
with initial conditions
\beqa
b_\pm(0)=1, \dot{b}_\pm(0)=0,
\nonumber\\
\alpha_\pm(0)=\dot{\alpha}_\pm(0)=0,
\eeqa
and compute easily the final energies with equation (\ref{EnergyNM}).
The values of the parameters are varied with a subroutine that minimises the sum of the final mode energies~(\ref{EnergyNM}) (we use the MatLab `fminsearch' and $n=0$ but note that the optimal final values of $b(t_f)$, $\alpha(t_f)$  and their derivatives would minimise the energies for any $n$).
The excess energy found with the optimal parameters for the normal modes is negligible in the range of final times depicted in figure \ref{fig7_2}.
%
%fix the free parameters so that they satisfy the boundary conditions that leave the normal modes excitationless~\ref{boundary1_2ion},~\ref{boundary2_2ion}.  We proceeded by fixing some values for the free parameters $c_{3-6}$, solving the auxiliary equations numerically and then recursively repeating the process in a shooting method .
%
% % % % % % % % % % % % % % % % % % % % % % % % % % % % % % % % % % % % % % % % %
% % % % % % % % % % % % % % % % % % % % % % % % % % % % % % % % % % % % % % % %
\begin{figure}[t]
\begin{center}
\includegraphics[width = 0.6\linewidth]{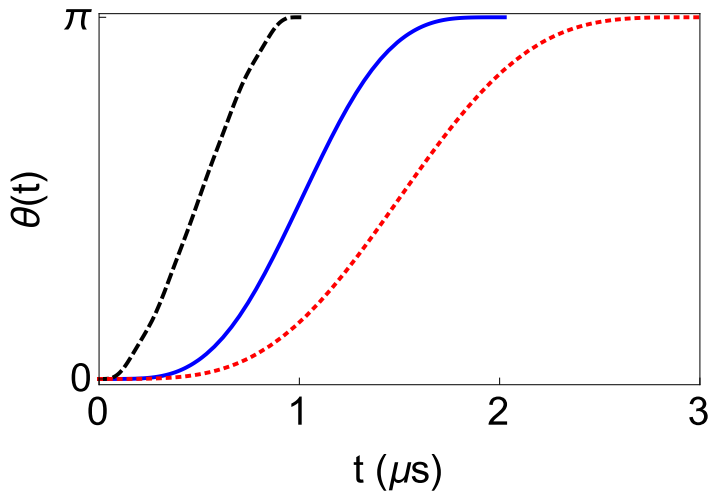}
\caption{\label{fig7_3}(Color online)
Evolution of the control parameter $\theta(t)$ for different final times when designed using all 4 free parameters.
Dashed black line:  $t_f=1$ $\mu$s, and optimisation parameters $c_{3-6}=(5.134,-5.360,59.577,91.234)\times 10^{-4}$;
Solid blue line:  $t_f=2$ $\mu$s, and optimisation parameters $c_{3-6}=(3.093,0.971,3.386,-6.036)\times 10^{-4}$;
Dotted red line:  $t_f=3$ $\mu$s, and optimisation parameters $c_{3-6}=(1.400,-0.270,0.182,-0.117)\times 10^{-4}$.
Other parameters as in figure \ref{fig7_2}.
}
\end{center}
\end{figure}
% % % % % % % % % % % % % % % % % % % % % % % % % % % % % % % % % % % % % % % % % % %
% % % % % % % % % % % % % % % % % % % % % % % % % % % % % % % % % % % % % % % % % % %
%
Once the free parameters are defined such that the design of $\theta$ minimises the excitation energy of the normal modes, we perform the quantum evolution driven by  the full Hamiltonian with (\ref{Hamiltonian_2ionrot}) to check the performance of the designed protocol. We use  the ``Split-Operator Method'', and the initial ground state is found performing an evolution in imaginary time.
Figure \ref{fig7_2} shows the final excitation, i.e., the excess energy with respect to the initial energy after performing the evolution with the full Hamiltonian (\ref{Hgeneral}) using  the potential~(\ref{Hamiltonian_2ionrot}). In figure \ref{fig7_2} (a) this excitation is depicted in a linear scale, and in figure \ref{fig7_2} (b) in a logarithmic scale. The results improve significantly by using more optimisation parameters. Even when using a single optimising parameter, the results are clearly better than the protocol without free parameters.
Figure \ref{fig7_3} shows some examples of the rotation protocols with 4 parameters for different rotation times.
%Finally, it can be said that the best of the optimising protocols (4 free parameters) reaches the threshold of 0.1 motional excitation quanta at a final time $t_f=1.05$ $\mu$s, whereas the adiabatic evolution reaches the same threshold at $t_f=2.23$ $\mu$s. That means that using our shortcut-to-adiabaticity protocol, one can accelerate the rotation of two ions by a factor of over 2.
%
%
% % % % % % % % % % % % % % % % % % % % % % % % % % % % % % % % % % % % % % % % %
% % % % % % % % % % % % % % % % % % % % % % % % % % % % % % % % % % % % % % % %
\begin{figure}[t]
\begin{center}
\includegraphics[width = 0.6\linewidth]{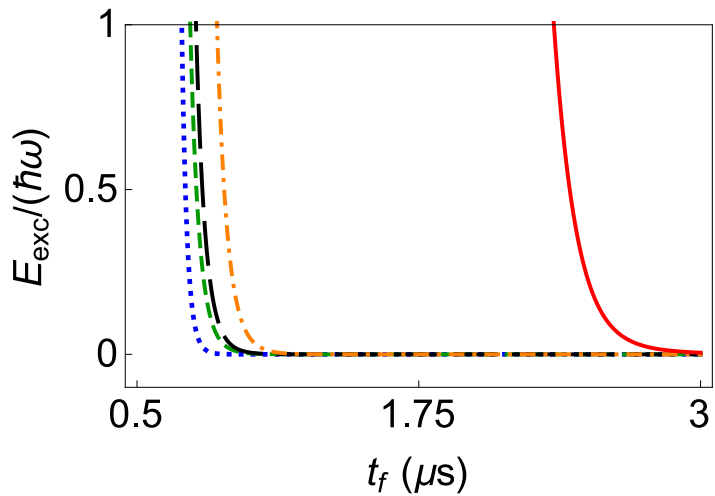}
\caption{\label{fig7_4}(Color online)
Two different ions. Exact energy excess (final minus initial energy) when the initial ground state is driven by the full potential~(\ref{Hamiltonian_2ionrot}) according to equation~(\ref{controlparameter}) for the parameters $c_{3-6}$ that minimise this excitation.
Dotted blue line:  protocol using all 4 free parameters in the ansatz for $\theta(t)$;
Short-dashed green line: only 3 free parameters, i.e., $c_6=0$;
Long-dashed black line: only 2 free parameters,  i.e., $c_5=c_6=0$;
Dash-dotted orange line: only one free parameter, i.e., $c_4=c_5=c_6=0$;
Solid red line fixes: $c_{3-6}=0$ so there is no optimisation.
The evolution was done for a $^{40}$Ca$^+$ and a $^9$Be$^+$ ion, with an external trap frequency for the Ca ion of $\omega_1/(2\pi)=1.41$ MHz and a total rotation angle $\theta_f=\pi$.
}
\end{center}
\end{figure}
% % % % % % % % % % % % % % % % % % % % % % % % % % % % % % % % % % % % % % % % % % %
% % % % % % % % % % % % % % % % % % % % % % % % % % % % % % % % % % % % % % % % % % %
%

%
%
%
\section{Two different ions}
%
%
%
%\subsection{Different ions: $d^3 k=\textrm{const}$}
%
Let us first explore some possible manipulations to make the modes separable when the ions are different.
From the expression of $\tan 2\mu$ in equation (\ref{decoupling_condh}), $d^3 k$ should be constant.
If the only parameter that depends on time is  $d$, this  condition  cannot be satisfied.
But if $k$ is allowed to be a time dependent controllable
parameter, it would be in principle possible.
If we set the constant as $B$ then, from equation (\ref{decoupling_condh}),  the relation
\begin{eqnarray}
  \tan 2\mu=\frac{4C_c\sqrt{m_1m_2}}{(m_1-m_2)(2C_c+B)},
 \end{eqnarray}
fixes $\mu$ to have independent dynamical modes. Using the expressions for $d$ and the $u_i$, this condition may be satisfied
for two values of $\dot\theta^2=a_lk$ for each $k$, $l=1,2$, where the $a_l$ are two constants.
%
%\begin{eqnarray}
% \left\{\begin{matrix}
%         \dot\theta^2&=&0,\\
%         \dot\theta^2&=&\frac{k}{2}\left(\frac{m_1+m_2}{m_1m_2}\right)
%        \end{matrix}\right.
%\end{eqnarray}
%
The proportionality between $\dot{\theta}^2$ and $k$, however, is problematic.
If we wish to approach $\dot\theta=0$
smoothly at the time boundaries, then $k\to 0$ there, which implies a vanishing trapping potential and $d\to\infty$.
%Alternatively sudden $\dot\theta$
A way out is explored in Appendix B making use of a more complex external trap potential with linear and quartic terms added,
as in reference \cite{Sagesser2020}.
In the main text we stay within the harmonic trap configuration with constant $k$ and renounce to separate the
modes. Thus a different, pragmatic strategy is adopted, minimising the excitation energy directly to find the rotation protocol.
%
%
%
%
%In the scheme of ion chains that have to be reordered, tackling the rotation of two different ions is of even more interest that the rotation of equal ions. However, a shortcut cannot be designed using the same method because of the
%coupling.
%
%
%
% % % % % % % % % % % % % % % % % % % % % % % % % % % % % % % % % % % % % % % % %
% % % % % % % % % % % % % % % % % % % % % % % % % % % % % % % % % % % % % % % %
\begin{figure}[t]
\begin{center}
\includegraphics[width=0.6 \linewidth]{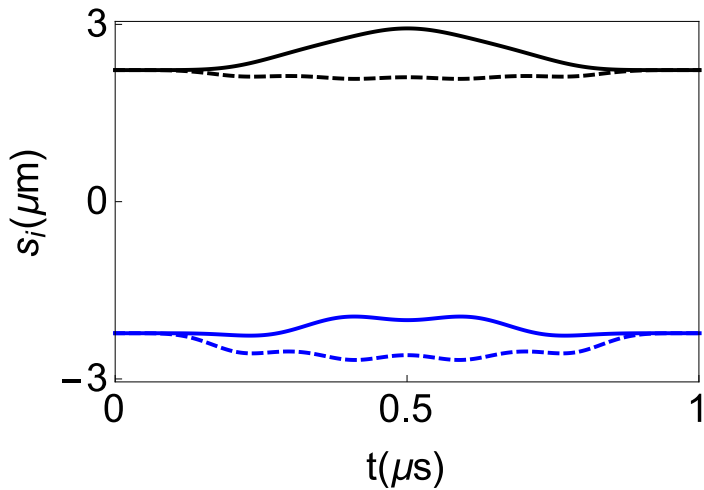}
%{fig7_5.eps}
\caption{\label{fig7_5}(Color online)
Equilibrium (dashed lines) and dynamical (solid lines) positions of the ions versus time:  $s_1^{(0)}$ and $s_1$ (Calcium ion, blue lines);  $s_2^{(0)}$ and $s_2$ (Berilium ion, black lines), for a final time $t_f=1$ $\mu$s and for the optimising parameters $c_{3-6}=(1.757,1.824,1.120,-0.234)\times 10^{-2}$ with the protocol in equation~(\ref{controlparameter}). The initial state is the ground state. The $s_i$ are average positions from the quantum dynamics.
}
\end{center}
\end{figure}
% % % % % % % % % % % % % % % % % % % % % % % % % % % % % % % % % % % % % % % % % % %
% % % % % % % % % % % % % % % % % % % % % % % % % % % % % % % % % % % % % % % % % % %
%
%That means that we cannot write it as two independent one dimensional harmonic oscillators, and thus the Lewis-Riesenfeld invariants \cite{Lewis1969} are of no use here \cite{Lizuain2017}. In Appendix \ref{dyndecoupling} we will attepmt adding a linear force to the potential to decouple the dynamics of the normal modes as inspired by \cite{Sagesser2020}.
%
%Here, instead, we use a brute force approach to optimise the protocol.
We use the same ansatz for the parameter control $\theta$ as in equation~(\ref{controlparameter})
%We try initially random values for the
%but instead of some differential equations as for the equal ion case, we
and solve the full (quantum) dynamics for the potential~(\ref{Hamiltonian_2ionrot}) to find the final excess energy for specific values of the free parameters $c_{3-6}$. Then, as in section \ref{equalionssec}, we minimise the excess energy letting the MATLAB subroutine  `fminsearch'
find the optimal parameters.

In figure \ref{fig7_4} we depict this final excitation, optimising the result using from 1 to 4 free parameters for the $\theta$, and compare it with the results for no free parameters. This direct minimisation provides even better results than the indirect one based on the normal mode energy in section \ref{equalionssec}. The best protocol (4 optimising parameters) gives an excitation below 0.1 quanta at a final time $t_f=0.56$ $\mu$s.
%, whereas for the protocol with no optimisation it is $t_f=2.6$ $\mu$s.
%That is, we get an improvement of almost a factor of 5.
The price to pay though, is that  the computational time required increases dramatically, as we have to solve the full dynamics of the system at every iteration of the shooting method we use to optimise, whereas in the method based on normal modes we only needed to solve four ordinary differential equations at each iteration. Figure \ref{fig7_5} shows the equilibrium and dynamical positions of both ions during the evolution for $t_f=1$ $\mu$s. The trajectories are not symmetric since the two ions experience different effective spring constants, see equation~(\ref{spco}).

\section{Discussion}

We have designed protocols to rotate a linear trap containing two ions, without final excitation.
%Reordering is one of the applications.
%Among other applications, this is a way
%to ordering of the ions without producing any additional excitation energy.
%Although the adiabatic protocols are not as slow as for other dynamical processes that are used in trapped ions, such as transport \cite{Palmero2014} or ion sepaNormal mode excitation
% ration \cite{Palmero2015}, we were still able to show a clear improvement in the final time required by using shortcuts to adiabaticity.
For two equal ions in a rotating, rigid harmonic trap, there are uncoupled dynamical normal modes. The separation facilitates inverse engineering since it is only necessary to  solve ordinary differential  equations for independent variables to minimise the final energy.
These Ermakov and Newton equations are for the auxiliary functions in the invariants associated with the uncoupled Hamiltonians.
Following this method and for a given ansatz for the rotation angle and for some allowed final excitation threshold, process-time lower limits are met due to the eventual failure of the small oscillation regime for very rapid rotations.
Faster processes can be achieved by increasing the number of parameters in the ansatz.  For two different ions in a harmonic trap, this method is not possible as the modes are coupled for a rigid trap, or can be uncoupled for a non-rigid trap but only for
impractical boundary conditions for the trap.
Instead we used direct optimisation of the rotation ansatz parameters with the full Hamiltonian. This direct approach is efficient with respect to the lower time limits but the  computational effort  is much more demanding.

A natural extension of this work would be considering different boundary conditions, for example a final rotating trap with $\dot\theta(t_f)\ne 0$, as in reference \cite{Urban2019}, to transfer an angular momentum to the chain.
Another possible future extension would be adding noises and perturbations to make the protocols robust with respect to them. Finally, specific protocols could be designed to simultaneously rotate longer chains of ions, although it is possible to sequentially rotate them in groups of 2 using the protocols designed here.

\section*{Acknowledgements}
We thank Uli Poschinger for discussions on the early stages of this paper.
This work was supported by the Basque Country Government (Grant No. IT986-16), and by the Spanish Ministry of Science and Innovation through projects PGC2018-101355-B-I00 and PGC2018-095113-B-I00 (MCIU/AEI/FEDER,UE).
% This work was supported by
% the Basque Country Government (Grant No. IT472-10),
% Ministerio de Econom\' ia y Competitividad (Grant No. FIS2012-36673-C03-01),
% and the program UFI 11/55 of the Basque Country University.
% A. Tobalina acknowledges financial support from Spanish Government via
% PGC2018-095113-B-I00 (MCIU/AEI/FEDER, UE)
%M.P. acknowledges a fellowship by UPV/EHU.
%
%

%\bibliographystyle{unsrt}
\bibliographystyle{iopart-num}
\bibliography{Biblio}

\appendix

\section{Magnetic force vs electric force\label{magnetic}}
Two charged particles moving in a direction perpendicular to the direction in which they are aligned experience a magnetic force, with magnitude
%. The force given by this interaction is defined as
%
\beq
F_{mag}=\frac{\mu_0}{4\pi}\frac{e^2}{r^2}|\vec{v}_1\times (\vec{v}_2\times\hat{r})|,
\eeq
where
%being $\times$ a vectorial product,
$\mu_0$ is the permeability constant, $\vec{v}_i$ the velocity vectors of each ion, $\vec{r}=\vec{s}_2-\vec{s}_1$ the position vector of ion 2 with reference to ion 1, and $\hat{r}=\vec{r}/r$. The Coulomb interaction, which is the only one considered so far, gives a force of magnitude
\beq
F_{el}=\frac{C_c}{{r}^2}.
\eeq
The ratio of these two forces is, using $\mu_0\epsilon_0=c^{-2}$, where $c$ is the speed of light,
\beq
R=\frac{F_{mag}}{F_{el}}=\frac{ |\vec{v}_1\times (\vec{v}_2\times\hat{r})| }{ c^2 }.
\eeq
With  $|\vec{v}_1|,|\vec{v}_2|\approx \frac{r}{2}\dot{\theta}$
we get
\beq
R\approx \frac{r^2\dot{\theta}^2}{4c^2}.
\eeq
For the protocols designed in the main text, the maximum values during the simulations at the represented times are $\dot{\theta}_{max}=5\times10^6$ s$^{-1}$ and $r_{max}=5.5\times 10^{-6}$ m
so
%. So, roughly, the maximum ratio between electric and magnetic forces is $R_{max}\sim c^{-2}$. For these parameters, the dipole-dipole the
the
magnetic interaction is negligible with respect to the electric force.

\section{Rotation of two different species ions based on dynamical normal modes\label{dyndecoupling}}
\begin{figure}[t]
\begin{center}
\includegraphics[width=0.8\linewidth]{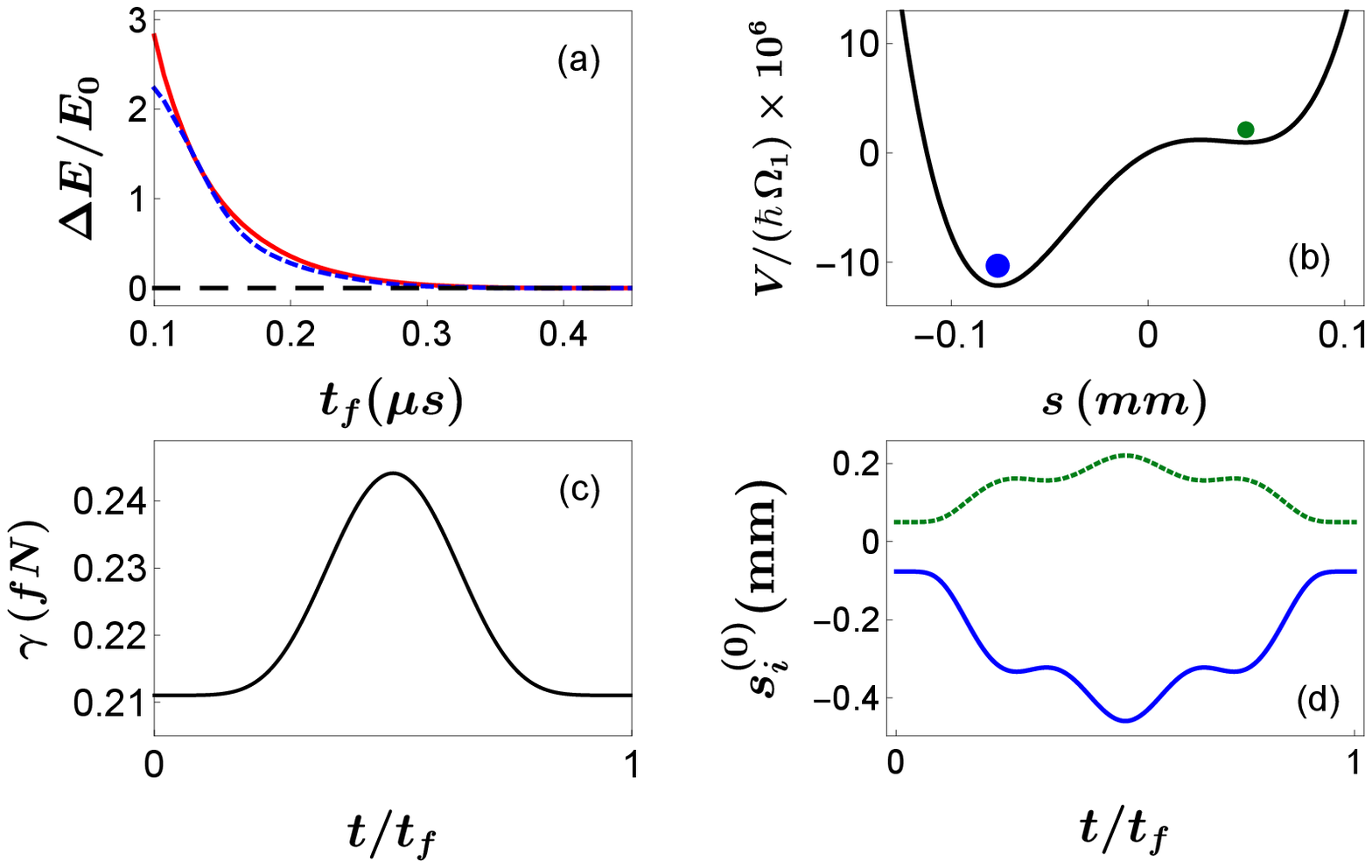}
\caption{
\label{Exc_gamma_V0}
(a) Normal mode excitation $\Delta E=E(t_f)-E(0)$ in units of the initial energy $E_0\equiv E(0)$ for different final times.
The protocol rotates a $^{40}$Ca$^+$ and a $^9$Be$^+$ ion in a double well potential with $m_1\omega_1^2=m_2\omega_2^2=-4.7$ pN/m and $\beta=0.52$ mN/m$^3$.
The initial state is a product of the ground states of each normal mode and thus, the energy of the system is computed as $E=E''_{0+}+E''_{0-}$, see equation~(\ref{EnergyNM}).
The solid red line represents a non-optimised protocol; blue dotted and black dashed and lines represent optimised protocols using one and two parameters respectively.
(b) Initial potential configuration and
(c) the required $\gamma(t)$, see equation~(\ref{dt}),
for the protocol with two optimisation parameters ($c_3=0.0059$ and $c_4=0.0285$) and $t_f=1$ $\mu$s.
(d) Corresponding evolution of the equilibrium positions, whose initial value is also represented in (b).
The blue solid line is for $^{40}$Ca$^+$ and the green dashed line for $^9$Be$^+$.
}
\end{center}
\end{figure}
If the matrix ${\bm{\mathsf{v}}}$ is time dependent, the normal modes get decoupled if ${\mathsf{v}}_{11}={\mathsf{v}}_{22}$,  see  equation (\ref{decoupling_cond}), i.e.,
\beq
\label{sepcon}
 \frac 1 m_1 \frac{\partial^2 V}{\partial s_1^2}\bigg|_{s_1^{(0)}} =\frac 1 m_2\frac{\partial^2 V}{\partial s_2^2}\bigg|_{s_2^{(0)}}.
\eeq
 %
%Thus, any potential that satisfies equation~\ref{sepcon} has dynamical normal modes that allow us to design the process using conventional STA techniques in each of them \cite{Lizuain2017,Sagesser2020}.
As explained in the main text, the rotation of different ions trapped by a rigid harmonic potential cannot be described in general in terms of dynamical normal modes. For a non-rigid one there is a formal solution which does not lead to practically useful boundary conditions.
Here we consider different confining potentials that obey equation~(\ref{sepcon}), and thus allow us to inverse engineer the rotation using the Lewis-Riesenfeld family of invariants.
%As in the main text $s_j^{(0)}$ are equilibrium positions and
%$d$ the distance between them.
We use for the equilibrium positions the parametrisation $s_1^{(0)}=s_0-d/2$ and $s_2^{(0)}=s_0+d/2$, where $s_0$ is the middle point between them.

Specifically we consider a tilted double well potential, which combines a repulsive harmonic potential with the confinement provided by the quartic term and a linear term \cite{Sagesser2020},
\begin{equation}
\label{quarticpot}
V = \gamma(t) (s_1+s_2) + \frac 1 2 u_1(t) s_1^2+\frac 1 2 u_2(t) s_2^2 + \beta (s_1^4+s_2^4)+ \frac {C_c}{s_2-s_1}.
\end{equation}
This gives the potential matrix
 \begin{equation}
 {\bm{\mathsf{v}}}=
  \left(\begin{array}{lr}
    \frac{\frac{2 C_c}{d^3}+u_1+12\left(-\frac{d}{2}+s_0\right)^{\!2}\beta}{m_1} & -\frac{2 C_c}{d^3\sqrt{m_1 m_2}}\\
    -\frac{2 C_c}{d^3\sqrt{m_1 m_2}} & \frac{\frac{2 C_c}{d^3}+u_2+12\left(\frac{d}{2}+s_0\right)^{\!2}\beta}{m_2}
  \end{array}\right).
\nonumber %
% \left(\begin{matrix}\tilde s_1-s_1^{(0)}\\ \tilde s_2-s_2^{(0)} \end{matrix}\right)
\end{equation}
The main-text equations from equation (\ref{effective}) to (\ref{omegas}) are still valid here.
We assume that the controllable parameters are the linear potential and the rotation speed. Equation~(\ref{sepcon}) is satisfied whenever $d$ obeys
\beqa
\label{eqd}
&\,&
%\frac1 {\beta  d}
\Big\{
A({m_1}-{m_2}) ({u_1}-{u_2})  +d^2\left[6 A \beta ({m_1} + m_2) + ({m_1}-{m_2}) ({u_1}-{u_2})^2\right]
\nonumber\\
&+& 24 \beta {C_c} d ({m_1}-{m_2}) +12 \beta ^2 d^6 ({m_1}-{m_2})
\Big\}=0,
\eeqa
where we have defined
\beq
A=\sqrt{d^3\! \left[24 \beta  {C_c}\!-\!12 \beta^{2}
   d^5\!-\!12 \beta  d^3 ({u_1}\!+\!{u_2})\!+\!d ({u_1}\!-\!{u_2})^2\right]}.
\eeq
The force $\gamma(t)$ that would produce the desired evolution for $d$ is
\beqa
\gamma &=& \frac 1 {108 \beta ^2 d^6} \Big\{18 \beta  {C_c} d^2 ({u_2}-{u_1})-24 \beta ^2 d^5 A - d ({u_1}-{u_2})^2 A
\nonumber\\
&-& 6 \beta  {C_c} A +36 \beta ^2 d^7 ({u_1}-{u_2}) + d^3 \left[-6 \beta A ( {u_1} + u_2)  -({u_1}-{u_2})^3\right] \Big\},
\label{dt}
\eeqa
and the corresponding evolution for the middle point between the ions is
\beq
s_0=\frac{A +d^2 ({u_1}-{u_2})}{12 \beta  d^3}.
\label{mpdw}
\eeq
The frequencies of the normal modes $\Omega_\pm$ can be analytically expressed in terms of $d$, the parameters that define the potential ($u_1$, $u_2$ and  $\beta$) and the masses $m_1$ and $m_2$, but they are too lengthy to be reproduced here. Provided that  equation~(\ref{eqd}) is satisfied, the rotation of the potential in equation~(\ref{quarticpot}) is governed by an uncoupled Hamiltonian of the form (\ref{18}), with the corresponding frequencies $\Omega_\pm$ and momentum shifts that read
\beq
p_{0\pm}=\frac{1}{\sqrt{2}} \left[(\sqrt{m_1} \pm \sqrt{m_2}) \dot s + (\sqrt{m_2} \mp \sqrt{m_1}) \frac{\dot d}{2} \right].
\eeq
%
%\beqa
%p_{0+}&=&\frac{1}{\sqrt{2}} \left[(\sqrt{m_1} + \sqrt{m_2}) \dot s_0 + (\sqrt{m_2} - \sqrt{m_1}) \frac{\dot d}{2} \right],
%\nonumber\\
%p_{0-}&=&\frac{1}{\sqrt{2}} \left[(\sqrt{m_2} - \sqrt{m_1}) \dot s_0 + (\sqrt{m_1} + \sqrt{m_2}) \frac{\dot d}{2} \right].
%\nonumber
%\eeqa

From here on the procedure to design the protocol is similar to the one explained in section \ref{equalionssec}. We start from the same ansatz for $\theta(t)$, see equation~(\ref{controlparameter}), which satisfies the boundary conditions~(\ref{cond1_2ion}) and (\ref{cond2_2ion}) by design, and search for the values of the free parameters that minimise the final excitation. Decoupling the dynamics of the system into independent dynamical normal modes, however, is more demanding here than for equal ions. We compute the necessary force, see equation~(\ref{dt}), and equilibrium positions, see Eqs.~(\ref{eqd}) and (\ref{mpdw}), for each test value of the free parameters in $\theta(t)$.
%
%\begin{figure}[t]
%\begin{center}
%\includegraphics[width= \linewidth]{fig_appendix}
%\caption{\label{Exc_doublewell} Normal mode excitation $\Delta E=E(t_f)-E(0)$ in units of the initial energy $E_0\equiv E(0)$ for different final times. The protocol rotates a $^{40}$Ca$^+$ and a $^9$Be$^+$ ion in a double well potential described $m_1\omega_1^2=m_2\omega_2^2=-4.7$ pN/m and $\beta=0.52$ mN/$\text{m}^3$. The initial state is a product of the ground states of each normal mode and thus, the energy of the system is computed as $E=E''_{0+}+E''_{0-}$, see equation~\ref{EnergyNM}.  Solid red line represents a non-optimised protocol. Black dashed and blue dotted lines represent optimised protocols using one and two free parameters respectively.}
%\end{center}
%\end{figure}
%
%
%We evaluate the outcome of our design based on the energy of the normal modes in Eq~\ref{EnergyNM}.

Figure \ref{Exc_gamma_V0}(a) shows that, for a rotation of a $^{40}$Ca$^+$ and a $^9$Be$^+$ ion chain, any of the protocols produce no excitations in the normal modes for processes as fast as $0.4$ $\mu$s.
%, a considerable improvement with respect to the rotation of the harmonic potential.
%Apparently, the implemented force, designed to decouple the dynamics of the normal modes, plays the role of a compensating force against the centrifugal force produced by the rotation and reduces the final excitation.
It also  illustrates the improvement of the results by increasing the number of free parameters for $\theta(t)$.
%using higher degree polynomial function to design the protocol.
Normal mode excitation is an approximation of the exact excitation, nevertheless, our results suggest that performing the rotation with the double well may provide excitationless protocols at short time scales.

Figure \ref{Exc_gamma_V0}(b) and (c) depict, respectively,  the initial potential and the required force $\gamma(t)$ for a specific rotation protocol using the tilted double well potential in equation~(\ref{quarticpot}). Notice that even the lowest value of the force,  at boundary times, produces a considerable bias with little to none barrier potential between the two wells. Despite this, each equilibrium position, whose evolution is depicted in figure \ref{Exc_gamma_V0}(c), initially lays in its own well. This unusual potential shape would be the price to pay for mode separability. We note that a potential bias may be imposed or cancelled using STA methods as well \cite{Martinez-Garaot2015}.

\end{document}